\documentclass[prb,twocolumn,amsmath,amssymb,superscriptaddress,floatfix]{revtex4}
\usepackage[english]{babel}
\usepackage{amsfonts}
\usepackage{graphicx}
\usepackage{times}
\usepackage{braket}
\usepackage{soul}
\usepackage{subcaption}

\usepackage{color}
\definecolor{nred} {RGB}{224,0,0}
\definecolor{nblue} {RGB}{28,130,185}
\definecolor{dgreen} {RGB}{78,138,21}

\newcommand{\mat}[1]{\mathrm{#1}}

\begin{document}

	\title{Light bipolarons in a system of electrons coupled to dispersive optical phonons}

	\author{K. \surname{Kova\v c} }
	\affiliation{J. Stefan Institute, 1000 Ljubljana, Slovenia}
	\affiliation{Faculty of Mathematics and Physics, University of Ljubljana, 1000
		Ljubljana, Slovenia}
	
	\author{J. \surname{Bon\v ca}}
	\affiliation{J. Stefan Institute, 1000 Ljubljana, Slovenia}
	\affiliation{Faculty of Mathematics and Physics, University of Ljubljana, 1000
		Ljubljana, Slovenia}

	\date{\today}
	\begin{abstract}
		
		We investigate the ground state properties of the bipolaron coupled  to  quantum dispersive optical phonons in the one-dimensional Holstein Hubbard model. We concentrate on the interplay between the phonon dispersion and the Coulomb repulsion and their mutual effect on the bipolaron effective mass, the binding energy, and the phase diagram. Most surprisingly, the sign of the curvature of the optical phonon dispersion plays a decisive role on the bipolaron binding energy in the presence of the  Coulomb repulsion $U$. In particular, when the sign of the phonon dispersion curvature matches the sign of the electron dispersion curvature, the bipolaron remains bound in the strong coupling limit even when  $U\to \infty$ and the binding emanates from the exchange of phonons between two electrons residing on adjacent sites. At moderate electron-phonon coupling a light bipolaron exists up to large values of $U$.  Finally, an intuitive explanation of the role of the phonon dispersion on the bipolaron binding energy is derived using the strong coupling limit where the binding emanates from the exchange of phonons between two electrons residing on adjacent sites which leads to enhanced stability of bipolarons at elevated Coulomb repulsion. 
	\end{abstract}  
	
	\maketitle
	
	\setcounter{figure}{0}
	
	\section{Introduction}
	
	The interaction between electrons and lattice vibrations, a phenomenon known as electron-phonon (EP) interaction, is a subject of extensive research in solid-state physics. This interaction significantly influences the physical properties of a diverse range of materials, including organic semiconductors \cite{Coropceanu2007, Chang2022, Fratini2020}, manganites \cite{Lanzara1998, Karmakar2022, Huang_2019}, and perovskites \cite{Miyata_2017, Cinquanta2019, Ghosh2020}. To gain a fundamental understanding of systems where EP interaction plays a crucial role, researchers have extensively investigated the Holstein model (HM) \cite{Holstein1959}. Although the HM is conceptually simple, it lacks an exact analytical solution. Consequently, researchers have employed a wide array of numerical methods to explore its static and dynamic properties. These methods include exact diagonalization techniques on finite lattices \cite{Alexandrov1994, Ranninger1992, Marsiglio1993, Fehske_1997, Fehske2000, Capone1997, Hohenadler2003, Fehske2007,marsiglio2022}, the density matrix renormalization group (DMRG) \cite{Jeckelmann1998, Zhang1998, Bursill1998}, diagrammatic approaches \cite{Ciuchi1997, Fratini2006, Prodanovic2019, Mitric2022}, with the momentum-averaged approximation \cite{Berciu2006, Goodvin2006, Adolphs2014, Adolphs2014_2, Carbone2021} being notably successful, as well as various Monte Carlo methods \cite{Prokofev1998, alexandrov1999, Cataudella2007, Assaad2008, Kornilovitch1998, Hohenadler2004, Spencer2005, Mishchenko2015, DeFilippis2015, Miladic2023},recently developed hierarchical equations of motion approach\cite{Vukmirovic2022},  and variational approaches \cite{Wellein1997, Wellein1998, Bonca1999, Ku2002, Barisic2002}.
	
	HM simplifies EP interactions by focusing on short-range interactions between charge carriers and lattice distortions, assuming that long-range interactions are effectively screened. The strength of this short-range coupling depends on the relative displacements between the atom hosting the charge carrier and its neighboring atoms. 
	There have been relatively few studies that explore short-range coupling to acoustic phonons \cite{Li2011, Hahn2021, Li_2013} since acoustic phonons primarily involve \textit{in-phase} motion of neighboring atoms, resulting in negligible relative displacements and weak electron-phonon coupling. As a result, HM typically disregards acoustic phonons. In contrast, optical phonons describe \textit{antiphase} atomic motion, leading to substantial relative displacements and, consequently, much stronger electron-phonon coupling. A common approximation is to include optical phonons in HM as dispersionless Einstein phonons.  This choice is motivated not only by the notion  that it provides a reasonable approximation when the phonon bandwidth is small compared to the average phonon frequency but also by the additional complexities that phonon dispersion introduces into both analytical and numerical treatments of the model. As a result, there are relatively few research papers \cite{Marchand2013, Bonca2021, Bonca2022, Jansen2022} dedicated to studying HM with dispersive optical phonons. These studies have revealed that phonon dispersion has a profound impact on polaron properties, affecting quantities such as the effective mass \cite{Marchand2013}, the optical conductivity \cite{Jansen2022}, and the spectral function \cite{Bonca2021, Bonca2022}. However, it is important to note that these works  primarily investigate the influence of dispersion on polaron properties and do not delve into the interactions between polarons.
	
	Our study focuses on the effect of phonon dispersion in a system consisting of two electrons  coupled to optical phonons. To account for the Coulomb interaction between charge carriers, we introduce a Hubbard term to the HM, resulting in the Holstein-Hubbard model (HHM). It is worth noting that HHM can exhibit different types of bipolarons depending on the values of its parameters, as demonstrated in previous research \cite{Bonca2000, Bonca2001}. In the regime where EP coupling dominates, both polarons tend to be localized on the same lattice site, leading to the formation of what is known as an $S0$ bipolaron. However, as the strength of the Coulomb repulsion increases, it becomes less favorable for polarons to occupy the same site. In this scenario, the bipolaron spreads out over multiple lattice sites, and its specific type is determined by the probability distribution of electron occupation. In the case  of a maximal probability for electrons to occupy neighboring sites, the  bipolaron is labeled  a $S1$ bipolaron. Similarly, for an $S2$ bipolaron, the probability is highest for electrons to occupy next-nearest neighboring sites, and so on. 
	
	We conjecture that phonon dispersion significantly impacts the stability of bound states when the phonon cloud spreads over multiple sites.  In systems with dispersion, phonons can facilitate interactions between electrons on different lattice sites, a possibility absent in systems lacking dispersion, which may in turn  lead to the bipolaron mass reduction. This observation is particularly interesting in the context of a prospective bipolaronic superconductivity based on the theory of the  Bose-Einstein condensation of weakly interracting bipolarons \cite{Alexandrov1981,alex_book_1995,Kabanov1996}. The challenge with bipolarons forming a superconducting state is that the superfluid transition temperature $T_c$ is inversely proportional to their effective mass. Historically, the effective mass of strongly bound bipolarons has been assumed to be large and to increase exponentially with EP coupling \cite{Chakraverty1998, Bonca2000, Macridin2004}, implying low values of $T_c$ in the context of bipolaronic superconductivity \cite{Chakraverty1998}. The degree to which the wavefunction of a bipolaron spreads out has a direct impact on its effective mass. For instance, the $S1$ bipolaron exhibits a significantly smaller effective mass compared to the $S0$ type\cite{Bonca2000}. However, with increased spreading of the wavefunction, the bipolaron binding energy diminishes.

	The idea of bipolaronic superconductivity has been recently revived, proposing a  model for phonon-mediated high-Tc superconductivity 
	where lattice distortions modulate the electron hopping amplitude that leads to small-size, yet light bipolarons that undergo Bose-Einstein condensation\cite{Zhang2023,sous2023}. Alternative research on a similar model, in contrast,  suggests the existence of a fragmented condensate of separated polaron pairs at elevated Coulomb repulsion.\cite{grundner2023}

	
	This paper is organised as follows. We first introduce the model and briefly  describe the method based on the efficient construction of the variational Hilbert space. In the results section we first present the phase diagram of the model containing two electrons with opposite spins.   A special emphasis is set on the asymmetry of the phase diagram with respect to the sign of the phonon dispersion. We proceed by an in-depth investigation of the spacial distribution of the bipolaron through  computation of the density-density correlation function. We next focus on the effect of the phonon dispersion and the Coulomb interaction on the binding energy and the effective mass. We study  the effect of the phonon dispersion from the perspective of a  prospective bipolaronic superconductivity. We conclude by deriving the bipolaron binding energy in the strong EP  coupling regime and in the limit of large Coulomb repulsion.

	\section{Model and method}
	
	\subsection{Model}
	
	We consider a system of two-electrons coupled to dispersive optical phonons described by the following Hamiltonian
	\begin{eqnarray} \label{hol}
		\vspace*{-0.0cm}
		\mathcal{H}_0 &=& -t_\mathrm{el}\sum_{j,s}(c^\dagger_{j,s} c_{j+1,s} +\mathrm{H.c.}) + {g} \sum_{j} \hat n_{j}(b_{j}^\dagger + b_{j})+ \nonumber \\
		&+& \omega_0\sum_{j} b_{j}^\dagger  b_{j} + t_\mathrm{ph}\sum_{j}(b^\dagger_{j} b_{j+1} +\mathrm{H.c.}) + \nonumber \\ &+& U\sum_{j}n_{j,\uparrow}n_{j,\downarrow}, 
		\label{ham}
	\end{eqnarray}
	where $c^\dagger_{j,s}$ and $b^\dagger_{j}$ are electron and phonon creation operators at site $j$ and spin $s$, respectively,  $\hat n_{j} = \sum_s c^\dagger_{j,s}c_{j,s}$ represents the  electron density operator and $t_\mathrm{el}$ the nearest-neighbor electron hopping amplitude. From here and on we set $t_\mat{el}=1$. The dispersive optical phonon band, represented as $\omega(q) = \omega_0 + 2t_\mathrm{ph}\cos(q)$, can be characterized by two parameters: $\omega_0$, which determines the central position of the band, and $t_\mathrm{ph}$, which governs its bandwidth. Note that when $t_\mat{ph}<0$, the signs of the phonon and the electron dispersion curvatures overlap.  The second term in  Eq.~(\ref{ham}) describes the interaction  between electrons and phonons, and the last term the on--site Coulomb repulsion. In the study of HM, it is customary to introduce a dimensionless parameter that characterizes the system, namely, the effective EP coupling strength denoted as $\lambda$. It is defined as $\lambda = \varepsilon_p/2t_\mathrm{el} = g^2/2t_\mathrm{el}\sqrt{\omega_0^2 - 4t_\mathrm{ph}^2}$, where $\varepsilon_p$ represents the polaron energy in the atomic limit ($t_\mathrm{el} = 0$) \cite{Marchand2013}.  In this work, we focus on the influence of the optical phonon dispersion on the system's behavior. We will express our findings in terms of the EP coupling  $g$ rather than $\lambda$, which  depends  on $t_\mathrm{ph}$.

	\subsection{Method}
	
	We have used the numerical method described in detail in
	Refs. \cite{Bonca1999,Ku2002,Bonca2021}. A variational subspace is constructed iteratively beginning with an initial state where both electrons are on the same site with no phonons. The subspace is defined on an infinite one-dimensional lattice.   The variational Hilbert space is then generated by applying  a sum of two off--diagonal operators:$\sum_{j,s}(c^\dagger_{j,s} c_{j+1,s} +\mathrm{H.c.}) + \sum_{j} \hat n_{j} (b_{j}^\dagger + b_{j})$, $N_h$ times taking into account the full translational symmetry. The obtained subspace is restricted in a sense that it allows only a finite maximal distance of a phonon quanta from the doubly occupied site, $L_\mathrm{max_1}=(N_\mathrm{h}-1)/2$, a maximal distance between two electrons $L_\mathrm{max_2}=N_\mathrm{h}$, and a maximal amount of phonon quanta at the doubly occupied site $N_\mathrm{phmax}=N_\mathrm{h}$, while on the site, which is $L$ sites away from the doubly occupied one, it is reduced to $N_\mathrm{phmax}=N_\mathrm{h} - 2L-1$. 
	We have used a standard Lanczos procedure \cite{Lanczos1950} to obtain static properties of the model. We analyse  the convergence of the method vs. the size of the variational Hilbert space as set by $N_h$  in Appendix B.

	\section{Results}
	
\begin{figure}[!tbh]
		\includegraphics[width=0.9\columnwidth]{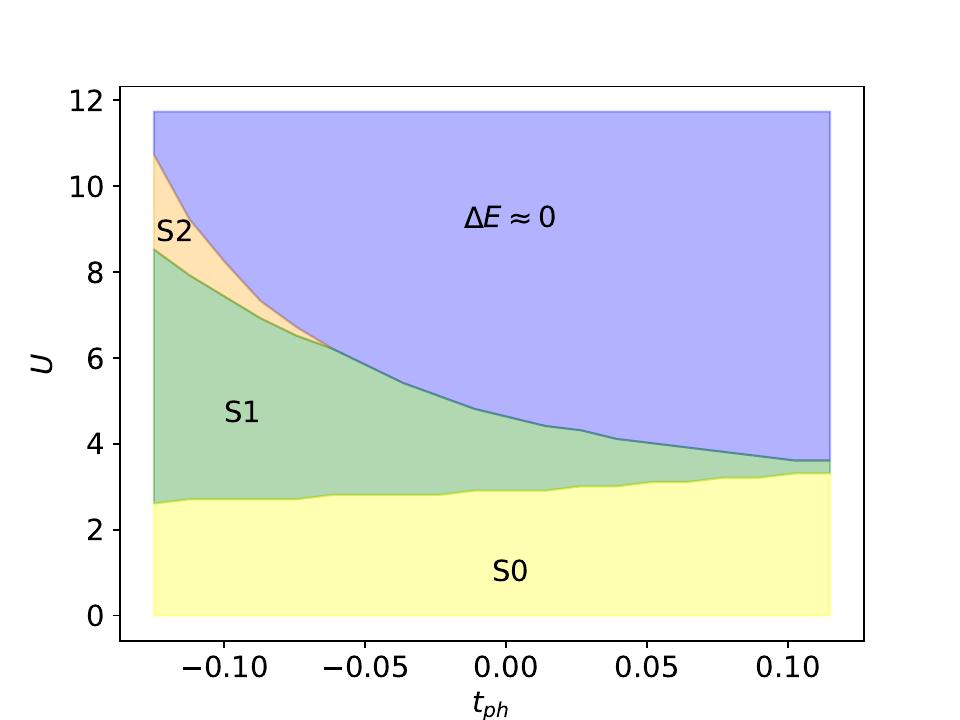}
		\captionsetup{justification=raggedright,singlelinecheck=false}
		\caption{Phase diagram ($U$, $t_\mathrm{ph}$) of our system of interest in the strong coupling regime, $g = \sqrt{2}$ and $\omega_0=t_\mat{el}=1$. There are four regions each represented by its own color. In three of them binding of polarons is energetically favorable, forming different regimes  of bipolaronic state denoted with labels $S0$, $S1$ and $S2$. In the blue regime the  Coulomb repulsion is strong enough to prevent the binding. In this region the  binding energy $\Delta E$ should be zero, but due to finite size effects it remains small but $\Delta E\gtrsim 0 $. For detailed analysis of finite size effects see Appendix B. In this and in all subsequent figures, we have used $N_h = 18$.}
		\label{fig1}
	\end{figure}

	The main result of our paper is a surprisingly large effect of phonon dispersion on static properties of bipolaron. We open this section with the phase diagram of our system of interest, Fig. \ref{fig1}, showing regions representing  different bipolaronic regimes depending on the strength of Coulomb repulsion $U$ and on the phonon dispersion $t_\mathrm{ph}$. We define the binding energy, $\Delta E = E_2 - 2E_1$, where $E_1$ and $E_2$ are ground state energies of one and two-electon systems, respectively. Different regimes  of bipolarons were determined by calculating the expectation value of density--density operator defined as 
	\begin{equation}
		\hat p(j)=\sum_i  \hat n_{i,\uparrow} \hat n_{i+j,\downarrow}.
		\label{gj}
	\end{equation}
	The calculated expectation value is connected with the probability for the electrons to be at the specific distance from each other. For example: $S0$ represent a bipolaron where the probability of electrons that occupy the same site is maximal.  
	 \\
	In the yellow, green and orange region electron-phonon coupling is strong enough to overcome Coulomb repulsion, leading to the formation of bipolarons. Within the yellow region, the ground state corresponds to the $S0$ bipolaron, while in the green and orange region, it is associated with the $S1$ and $S2$ bipolarons. The blue region represents separate   polarons where $\Delta E \approx 0$. 
	
	The most important result, observed in Fig. \ref{fig1},  is the apparent asymmetry of the phase diagram with regard to the sign of $t_\mat{ph}$. It seems that the bipolaron remains stable up to larger values of $U$ when $t_\mat{ph}<0$. While the stability of the $S0$ bipolaron shows weak dependence on $t_\mat{ph},$ the larger  $S1$ and $S2$ bipolarons remain stable at much larger values of $U$.  For example, at moderrate EP coupling $g=\sqrt{2}$ and  $t_\mat{ph}\sim -0.125$, with increasing $U$ the radius of the bipolaron increases to avoid the strong Coulomb repulsion as it crosses  over from the $S0$ via $S1$ towards $S2$ until at $U\sim 10$ the bipolaron separates into two polarons.  In the Appendix A  we show the phase diagram $(U,g)$ at fixed $t_\mat{ph}<0$. In  Section IV we show that in the  strong coupling limit the bipolaron remains bound even for $U\to \infty$ and the binding energy is up to the leading order in $g$ given by
	\begin{equation} \label{delta}
		\Delta E\sim
		\begin{cases}
			2t_\mat{ph}g^2/\omega_0^2; & t_\mat{ph}\lesssim 0\\
			0; & t_\mat{ph}\geq0.
		\end{cases}
	\end{equation}
	
	\begin{figure}[!tbh]
		\includegraphics[width=0.9\columnwidth]{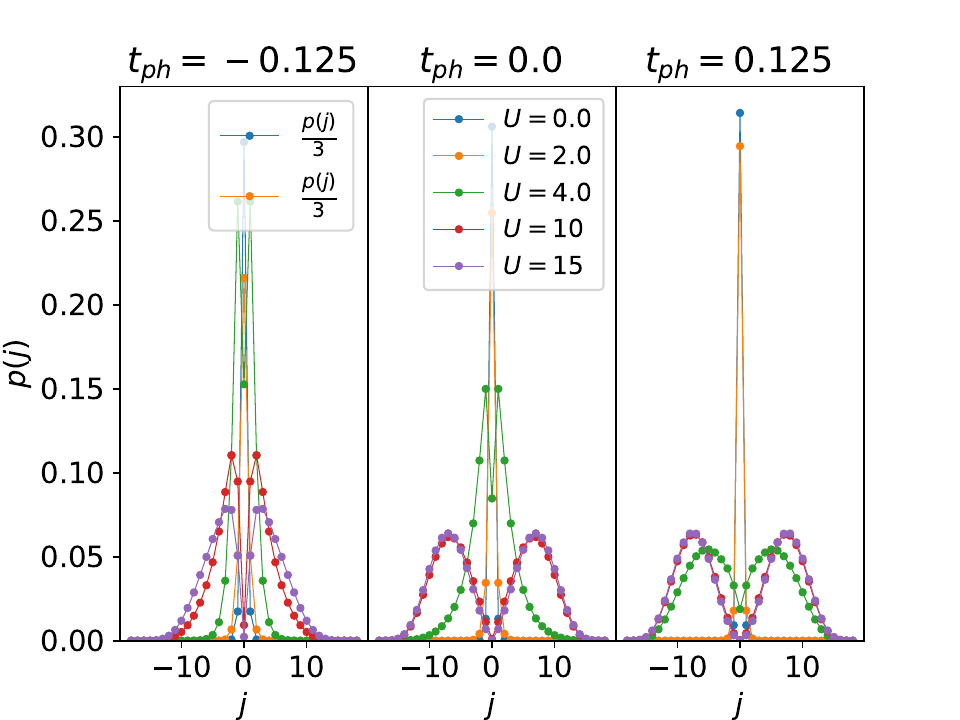}
		\captionsetup{justification=raggedright,singlelinecheck=false}
		\caption{The expectation value of density--density operator $p(j)$ in the ground state of the  two-electron system and   $\omega_0=1$, $g=\sqrt{2}$.  We have used   selected specific values of the on-site Coulomb interaction parameter $U$.  These particular choices of $U$ were made to illustrate various bipolaronic regimes present in our system. To enhance clarity in our presentation, we scaled $p(j)$ by a factor of 3 in cases where the $S0$ bipolaron is present. This scaling compensates for the pronounced peak in $p(j)$ at $j=0$ associated with the $S0$ bipolaron state. }
		\label{fig2}
	\end{figure}

	In  Fig. \ref{fig2} we present $p(j)$ for selected values of $U$ in the system with upward phonon dispersion ($t_\mathrm{ph} = -0.125$), without a dispersion ($t_\mathrm{ph} = 0.0$) and with downward phonon dispersion ($t_\mathrm{ph} = 0.125$). For clarity values of $p(j)$ are divided by 3 for the  $S0$ bipolaron, that is at $U = 0.0, 2.0$. In the $S0$ regime $p(j)$ is sharply peaked at $j=0$.  By increasing $U$ the peak at $j=0$ diminishes and broadens. Green lines in the  left and the central graph correspond to $S1$ bipolaron. We observe a pronounced dip at $j = 0$ followed by a peak at $j=\pm 1$ and an exponential decrease of $p(j)$ at larger distances, consistent with a bound state. 
	In case of $t_\mathrm{ph} = -0.125$ and $U = 10$,  $p(j)$  reaches its maximum value at $j = \pm 2$.  This state is designated   the  $S2$ bipolaron. For even larger values of $U$, the ground state no longer supports bound polarons. It is interesting to note that the transition to the unbound state (consisting of two separate polarons) requires a stronger Coulomb interaction as $t_\mathrm{ph}$ decreases. 
	In the unbound state (see results for  $t_\mat{ph}=0.0,0.125$ and $U=10,15$), $p(j)$ exhibits a distinctive behavior: it initially increases with increasing  $|j|$, it reaches a maximum around $|j|\sim 8$, and approaches  zero at even larger $|j|$.  
	This behavior is a consequence of the limited  variational subspace construction.  In the full Hilbert space, $p(j)$ would monotonically increase with $j$ while obeying the sum--rule $\sum_j p(j)=1$, reflecting the polaron's preference for maximizing their separation. However, our variational subspace allows only states up to  a maximal distance between polarons $L_\mathrm{max_2}$.
	Consequently, $p(j)$ reaches its maximum at a finite value of $j < L_\mathrm{max_2}$.
	\begin{figure}[!tbh]
		\begin{subfigure}[b]{0.9\columnwidth}
			\centering
			\includegraphics[width=1\linewidth]{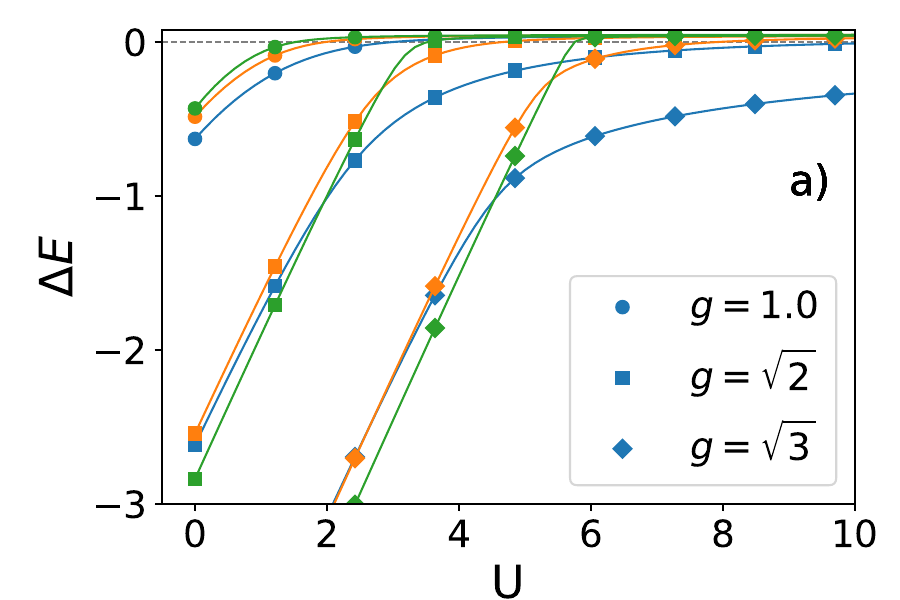} 
			\phantomsubcaption{}
			\label{fig3a}
		\end{subfigure}
		
		\begin{subfigure}[b]{0.9\columnwidth}
			\centering
			\includegraphics[width=1\linewidth]{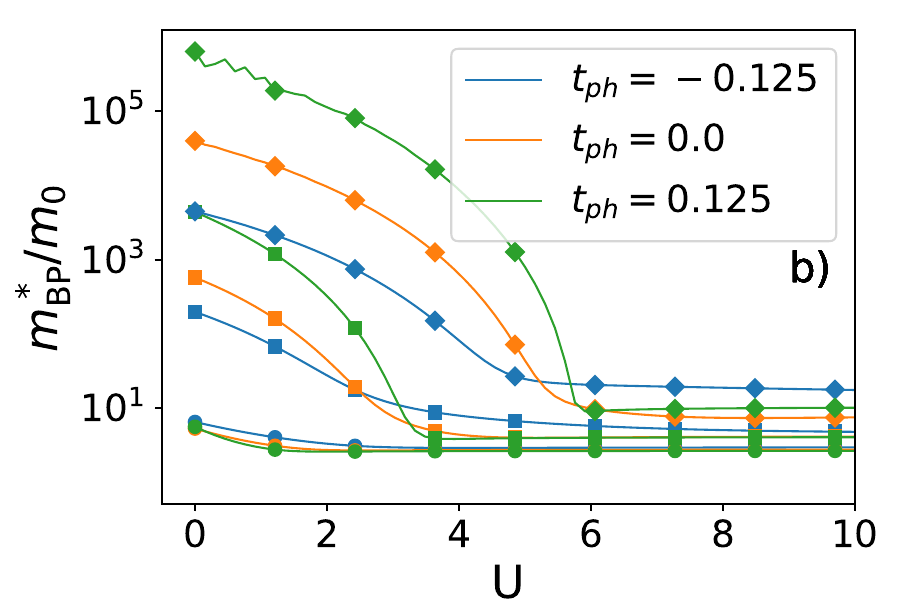} 
			\phantomsubcaption{}
			\label{fig3b}
		\end{subfigure}
		\captionsetup{justification=raggedright,singlelinecheck=false}
		\caption{Binding energy and effective mass are plotted as functions of the parameter $U$ in a system characterized by $g$ belonging to the set: $\left[1.0, \sqrt{2}, \sqrt{3}\right ]$ and $t_\mathrm{ph}$ drawn from the set: $\left [-0.125, 0.0, 0.125\right ]$, and $\omega_0=1$. To improve the visualization, only every $8^\mat{th}$ data point is shown on the plot. 
			We observe a non-monotonic behavior of effective mass at small $U$ for the case of $g = \sqrt{3}$ and $t_\mathrm{ph} = 0.125$. Due to a large effective mass, the bipolaron is nearly localized; consequently, the ground state energy is nearly independent of the wavevector, which in turn renders the effective mass ill-defined.}
		\label{fig3}
	\end{figure} 
	
	Strong EP interaction with Einstein optical phonons is responsible for the formation of bipolarons with an excessive large effective mass that increases exponentially with $g$, see  Refs.\cite{alex_book_1995,alexandrov1999,Bonca2001,Bonca2000}.  An important challenge is to search for a physically relevant microscopic EP coupled model that would allow for a strongly bound bipolaron with a small effective mass. Since the introduction of the phonon dispersion represents a physically relevant generalization of the standard HM, we now explore the effect of $t_\mat{ph}$ on the binding energy and the effective bipolaron mass. In accordance with the approach introduced in \cite{Bonca1999}, we define the effective bipolaron mass $m^*_{\mathrm{BP}}$, as
	\begin{equation}
		\frac{m^*_{\mathrm{BP}}}{m_0} = 2t_\mathrm{el}\bigg[\frac{\partial^2E_2(k)}{\partial k^2}\bigg]_{k=0}^{-1}.
	\end{equation}
	The effective mass is expressed in units of the effective mass of a free electron, defined as $m_0 = 1/2t_\mathrm{el}$. The second derivative of the ground state bipolaron energy dispersion, denoted as $E_2(k)$, is calculated using finite differences in the vicinity of $k=0$. 
	
	Fig. \ref{fig3} shows calculated effective mass and binding energy as a function of $U$ for different values of $g$ and $t_\mathrm{ph}$. At small $U$ the system is in the regime of $S0$ bipolaron characterized by exceptionally large effective mass and binding energy. The effective mass experiences a substantial reduction when the Coulomb repulsion becomes sufficient to prevent double occupancy.  Simultaneously, the absolute value of $\Delta E$ decreases as Coulomb repulsion counters the tendency of polarons to form bound states. However, as depicted in Fig. \ref{fig3a}, in a strongly coupled system with negative $t_\mathrm{ph}$, the binding energy exhibits an elongated tail, gradually approaching zero, but retaining a substantial value even as the effective mass remains of the order $m^*_\mat{BP}/m_0\sim 10$ (Fig. \ref{fig3b}). 
	
	We also observe a more gradual decrease of $m^*_\mat{BP}$  with increasing $U$ for  $t_\mat{ph}<0$ in contrast to its steeper descent at  $t_\mat{ph}>0$.   For  $t_\mat{ph}<0$, the bipolaron with increasing $U$  evolves from the S0 bipolaron towards the region of a stable S1 bipolaron, which in turn shrinks with increasing $t_\mat{ph}>0$  as observed in Fig.~\ref{fig1}.    In the strong EP coupling regime, the effective mass of the S0 bipolaron scales as $m_\mat{BS0} \sim \exp[4(g/\omega_0)^2]$ while the effective mass of the S1 bipolaron is in comparison much smaller and scales as $m_\mat{BS1} \sim \exp[(g/\omega_0)^2]$, see Ref.\cite{Bonca2000}.

	
	This observation looks promising  in the framework of recent theories based on the bipolaronic superconductivity \cite{Zhang2023} as it suggests the emergence of relatively strongly bound light bipolarons at negative values of $t_\mathrm{ph}$. However, it is crucial to consider that the temperature at which bipolarons undergo a superfluid transition  $T_c$  is influenced not only by their effective mass but also by their density, as noted in previous studies \cite{Pilati2008, Zhang2023}. In the parameter space where the effective mass is  small, e.g. $m^*_\mathrm{BP}/m_0\lesssim 10$, within  $S1$ and  $S2$ in Fig.~\ref{fig1},  $p(j)$ is distributed across multiple sites. We determined $T_c$   using the formula valid for the  superfluid transition  of a gas of hard-core bipolarons in 2D from\cite{Pilati2008, Zhang2023}
	\begin{equation} \label{Tc}
		T_c \approx
		\begin{cases}
			\frac{0.5}{m^*_\mathrm{BP}R_\mathrm{BP}^2}; & R_\mathrm{BP}^2 \geq 1, \\
			\frac{0.5}{m^*_\mathrm{BP}}; & \textrm{otherwise}.
		\end{cases}
	\end{equation}
	Here, $R^2_\mathrm{BP}= \sum_{j}p(j)j^2$ and $m^*_\mathrm{BP}$ represent the average size and effective mass of a bipolaron, respectively.  A word of caution:  our calculation is based on the 1D system; nevertheless, the effective mass in the Holstein model only weakly depends on the dimensionality of the problem\cite{Ku2002}. Our naive estimate of  $T_c$ based on the 1D computation should be considered only as an attempt to predict  the effect of the phonon dispersion on $T_c$ in higher dimensions.   

\begin{figure}[!tbh]
		\includegraphics[width=0.9\columnwidth]{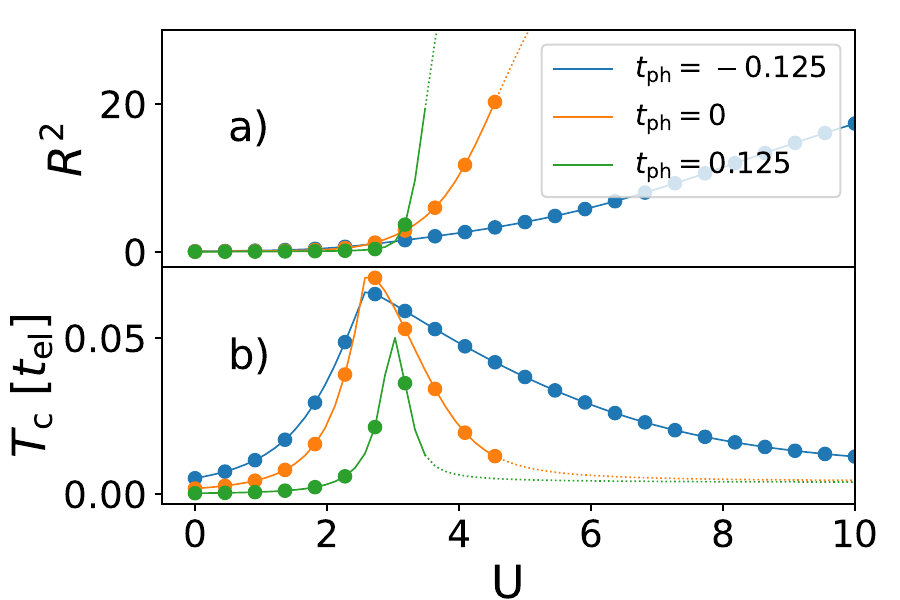}
		\captionsetup{justification=raggedright,singlelinecheck=false}
		\caption{The mean-square radius of bipolarons in a) and their corresponding transition temperatures in b), computed using Eq. \ref{Tc}, as functions of the parameter $U$ in a system with $g = \sqrt{2}, \omega_0=1$ and three distinct representative values of $t_\mathrm{ph}$. For improved clarity, only every $8^\mat{th}$  data point is displayed with a circle. These quantities are calculated over the entire selected range of $U$. Hence, dashed lines are employed to emphasize the unbound region where $R_\mathrm{BP}^2$ is expected to diverge in the thermodynamic limit.}
		\label{fig4}
	\end{figure}

	We conclude that dispersion does not substantially affect the maximum transition temperature at specific values of $U$. This observation arises from the fact that the maximum transition temperature is attained within the $S0$ regime, which exhibits minimal dependence on the parameter $t_\mathrm{ph}$. Nevertheless, the dispersion strongly affects the  range of $U$ values over which the transition occurs, as illustrated in Fig. \ref{fig4}.
	 \\
	By introducing a physically relevant dispersion term, we observed a reduction in the effective mass of a bound bipolaron, but only for  $t_\mat{ph}<0$. However, this change does not lead to a substantial increase in $T_c$. 
	Finally, we propose that the effective mass can be further decreased by expanding the range of interaction beyond just the on-site coupling. To explore this avenue,  we modify the Hamiltonian as follows 
	\begin{equation}
		\mathcal{H'} = \mathcal{H}_0 + V\sum_{i}\hat n_{i}\hat n_{i+1} + {g_2} \sum_{i,j=i\pm 1} \hat n_{i}(b_{j}^\dagger + b_{j}).
	\end{equation}
	This extension includes Coulomb interaction \cite{Lin1995} and electron-phonon coupling \cite{alex_book_1995,alexandrov1999} that extends to neighboring sites.  Consistent with the approach in \cite{alex_book_1995,alexandrov1999}, we set $g_2 = g / 8$. We arbitrarily decide for Coulomb interaction to exhibit a similar decay and set $V = U / 10.$
	\begin{figure}[!tbh]
		\begin{subfigure}[b]{0.5\columnwidth} 
			\centering
			\includegraphics[width=1\linewidth]{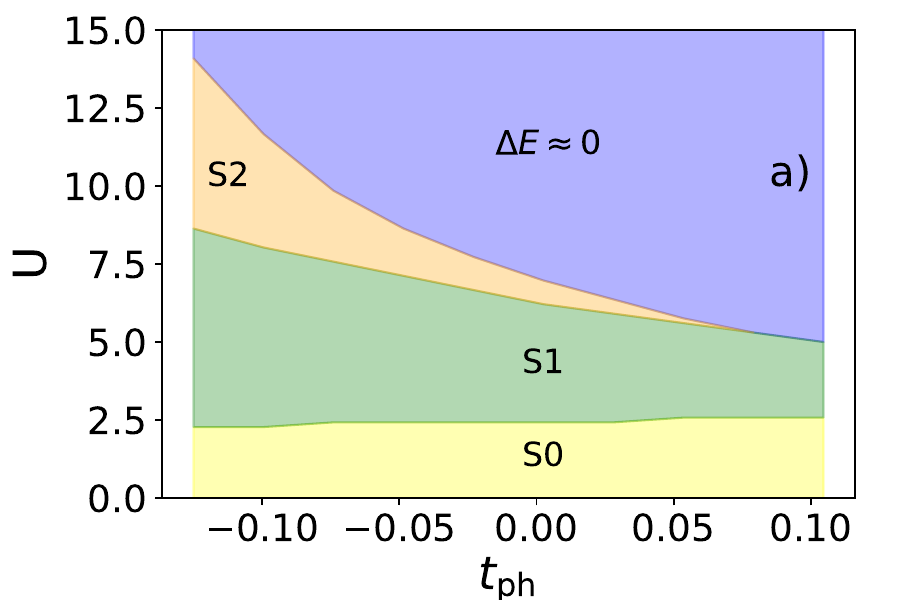} 
			\phantomsubcaption{}
			\label{fig5a}
		\end{subfigure}
		\begin{subfigure}[b]{0.5\columnwidth}
			\centering
			\includegraphics[width=1\linewidth]{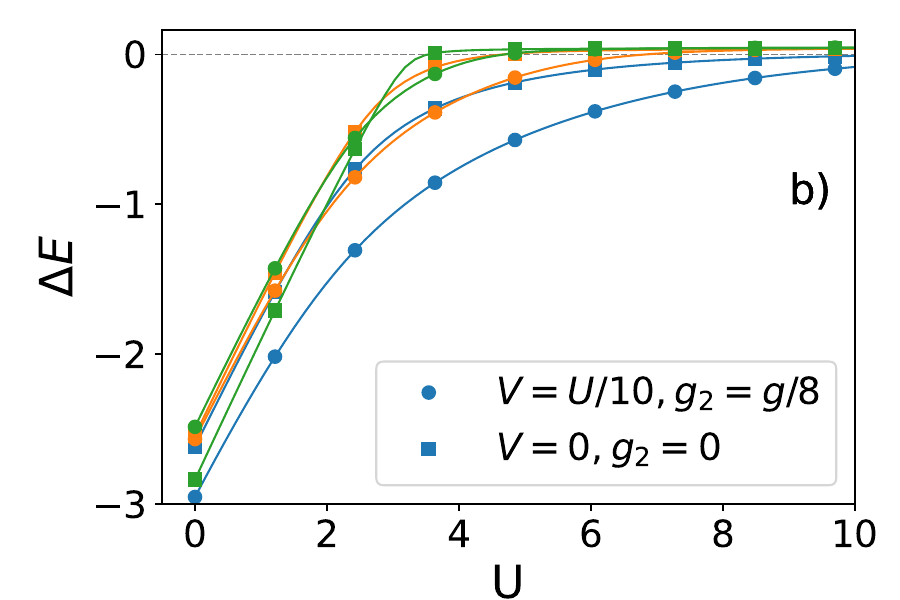} 
			\phantomsubcaption{}
			\label{fig5b}
		\end{subfigure}
		
		\begin{subfigure}[b]{0.5\columnwidth}
			\centering
			\includegraphics[width=1\linewidth]{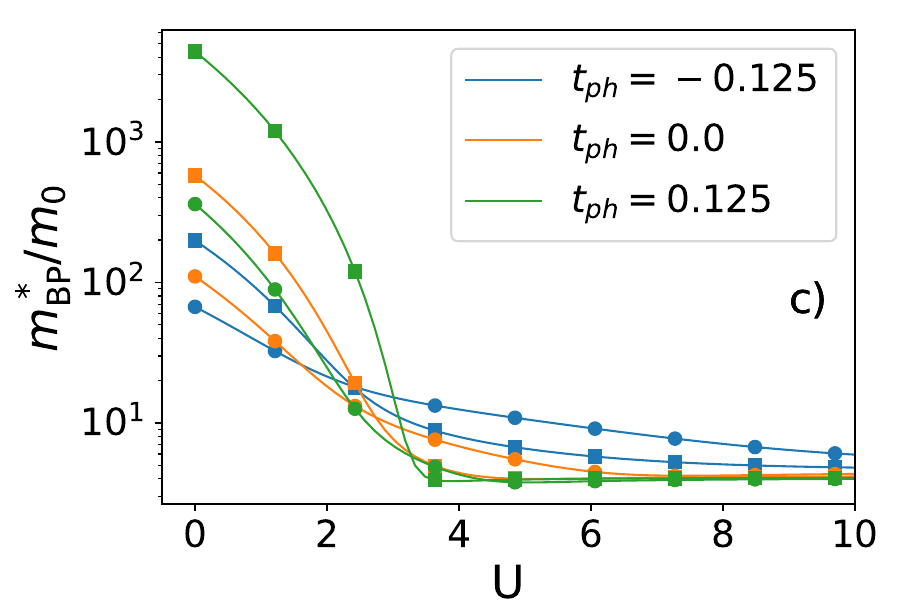} 
			\phantomsubcaption{}
			\label{fig5c}
		\end{subfigure}
		\begin{subfigure}[b]{0.5\columnwidth}
			\centering
			\includegraphics[width=1\linewidth]{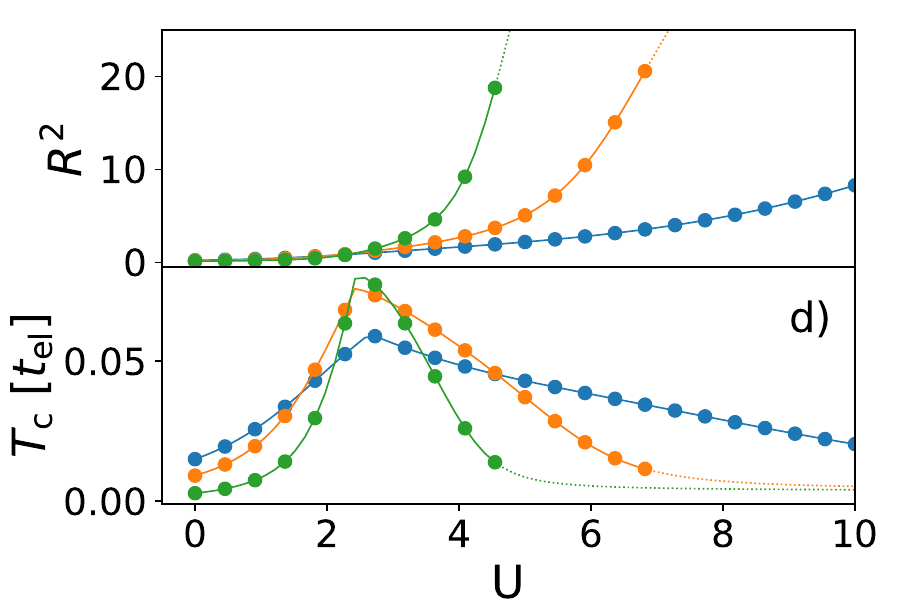} 
			\phantomsubcaption{}
			\label{fig5d}
		\end{subfigure}
		
		\captionsetup{justification=raggedright,singlelinecheck=false}
		\caption{a) The phase diagram ($U$, $t_\mathrm{ph}$) of a model with extended  interactions in the strong coupling regime ($g = \sqrt{2}, \omega_0=1$); b) comparison of $\Delta E$ between the Holstein (squares) and the extended Holstein models (circles) at $g = \sqrt{2}$ with three distinct  phonon dispersions; c) comparison of effective mass between the Holstein and extended Holstein models for the same parameters as in b);  d) the mean-square radius of bipolarons and their corresponding transition temperatures computed only for the model with extended interactions. 
		}
		\label{fig5}
	\end{figure} 
	
	As seen in  Fig.~\ref{fig5a}) and by comparison with  Fig.~\ref{fig1} , the extended interaction expands the range of $U$ where  $S1$ and $S2$ bipolarons exist. In addition, the $S2$ bipolaron regime extends even towards  $t_\mathrm{ph}>0$. In Figs. \ref{fig5b}) and \ref{fig5c}), which illustrate the impact of extended interaction on $\Delta E$ and $m^*_\mathrm{BP}$, respectively, we observe that extended interaction significantly reduces effective mass (by nearly an order  of magnitude) at small values of $U\lesssim 2$ , while having a minor effect on $\Delta E$. 
	This leads to broader and higher peaks in $T_c$ (Fig. \ref{fig5d}) in comparison with results presented in Fig. \ref{fig4}b).
	
	\section{Perturbational approaches}
	\subsection{Strong coupling regime}
	
	In order to explain the influence of the sign of the phonon dispersion $t_\mat{ph}$ on the stability of $S1$ and $S2$ bipolaron we turn to the strong EP coupling  and large-$U$ limit. In this case the state of two electrons governed by the Hamiltonian $\mathcal{H}_0$ can be described with a wavefunction $\psi_n$ where $n>0$ represents the distance between the two  particles and $\psi_0=0$ in the $U\to\infty$ limit. In the lowest order in the strong  EP coupling limit  the $\psi_n$ satisfies the following set of equations: 
	\begin{eqnarray}
		E_2(k) \psi_1 &=& (\epsilon_0 -2\tilde{g}^2 \omega_0) \psi_1 -t(1+e^{ik})\psi_2,\nonumber \\
		E_2(k) \psi_n &=& -2\tilde{g}^2 \omega_0\psi_n - t(1+e^{ik})\psi_{n+1} - \nonumber\\
		&-&t(1+e^{-ik})\psi_{n-1}; ~n>1,
		\label{str}
	\end{eqnarray}
	
	where $k$ is the total wavevector of the pair, $\epsilon_0 = 2t_\mat{ph}\tilde{g}^2$,  $t=t_\mat{el}e^{-\tilde{g}^2}$, and  $\tilde g = g/\omega_0$. Note that $\epsilon_0$ depends linearly on $t_\mat{ph}$ and  it is obtained from
	\begin{equation}
		\epsilon_0=\langle \tilde g_i \vert\langle\tilde g_{i+1}\vert  t_\mathrm{ph}\sum_{j}(b^\dagger_{j} b_{j+1} +\mathrm{H.c.}) \vert \tilde g_i \rangle\vert\tilde g_{i+1}\rangle,
	\end{equation}
	where $\vert \tilde g_i\rangle\vert \tilde g_{i+1}\rangle$ represent a product of coherent states $\vert \tilde g_i\rangle = \exp(-\tilde g^2/2 + \tilde g b^\dagger_i)c^\dagger_i\vert\emptyset\rangle$ of two  electrons on neighbouring sites.
	
	In case when $\epsilon_0<0$ one can expect that a bound state exists providing $\epsilon_0$ is less than a certain threshold value as shown below. Equations in Eq.~\ref{str} are equivalent to a system with one particle on a tight-binding semi--infinite chain with energy $\epsilon_0$ on the first site. The ansatz for a bound state of two electrons is given by:
	\begin{equation}
		\psi_n = \psi_2 e^{(-ik/2-\kappa)n};~~n\geq 2,
	\end{equation}
	and leads to 
	\begin{eqnarray}
		E_2(k)&=&-4t_\mat{el}e^{-\tilde{g}^2}\cos(k/2)\cosh{\kappa}-2\tilde{g}^2\omega_0,\\
		\kappa &=& \ln\left[ {-t_\mat{ph}\tilde{g}^2\over t_\mat{el}\exp{(-\tilde{g}^2)}\cos(k/2)}\right ].
	\end{eqnarray}
	The condition for a bound state at $k=0$ (the ground state is at $k=0$ providing $t_\mat{el}>0$)  is fulfilled when $\kappa>0$ that leads  $-t_\mat{ph}\tilde{g}^2 >  t_\mat{el}\exp{(-\tilde{g}^2)}$. The binding energy of a bipolaron is given by $\Delta E=E_2(k=0)-2E_1(k=0)$ where the single polaron energy in the strong coupling regime is $E_1(k=0)=-2t_\mat{el}e^{-\tilde{g}^2}-\tilde{g}^2\omega_0$, and 
	\begin{equation} 
		\Delta E= 2 t_\mat{ph}\tilde{g}^2 + 2t_\mat{el}e^{-\tilde{g}^2}\left [{t_\mat{el}e^{-\tilde{g}^2}\over t_\mat{ph}\tilde{g}^2} + 1\right ]. 
		\label{deltast}
	\end{equation}
	Note that the bipolaron is in the strong EP limit bound only when $t_\mat{ph}<0$.
	\begin{figure}[!tbh]
		\includegraphics[width=0.9\columnwidth]{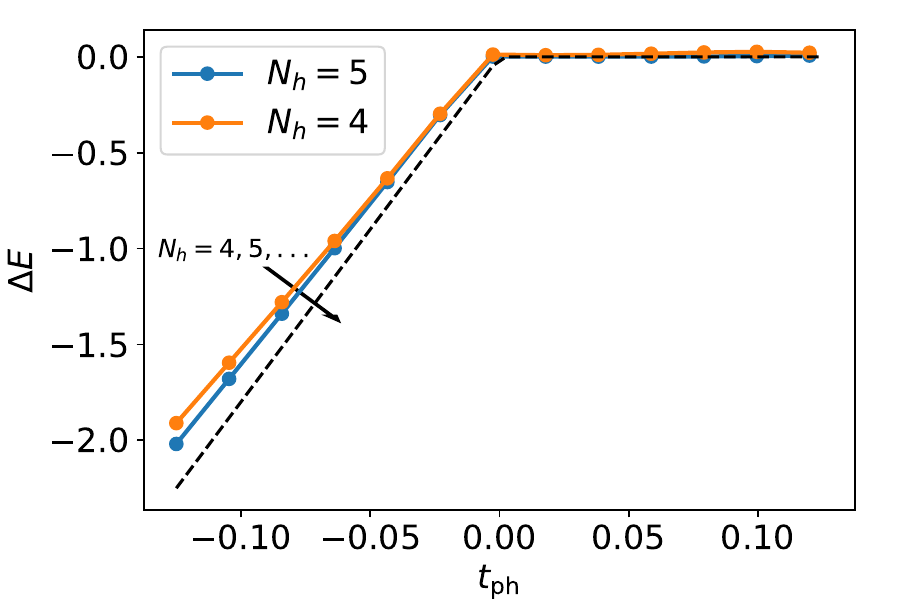}
		\captionsetup{justification=raggedright,singlelinecheck=false}
		\caption{Comparison of the analytical prediction (Eq.~\ref{deltast}) depicted by the  dashed line with numerically calculated $\Delta E$ for two values of $N_h$ and $g=3$.}
	\label{fig6}
\end{figure} 
In Fig.~\ref{fig6} we show comparison between the analytical result obtained from Eq.~\ref{deltast} and numerical results in the strong coupling regime, i.e. $g=3$ and large $U=10$.  The results demonstrate the convergence of the numerical calculations toward the analytical prediction as $N_h$ increases.  Since in the strong EP coupling regime the  electronic hopping is exponentially suppressed, numerically reliable results are obtained already on small lattice sizes, i.e. $L=N_h$. However, one needs to allow for a large maximal number of phonons in the system.  In this case we have used up to $N_\mathrm{phmax}=130$ phonons.

\subsection{Weak coupling regime}

\begin{figure}[!tbh]
	\includegraphics[width=1.0\columnwidth]{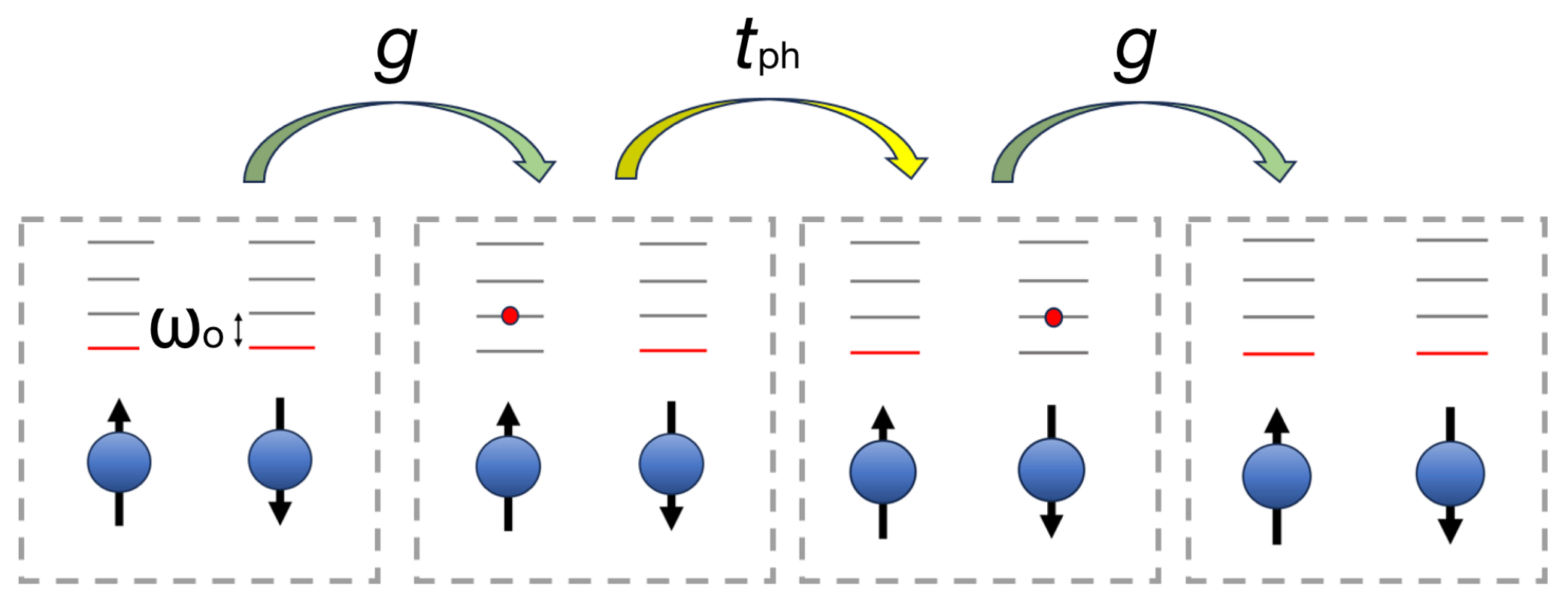}
	\captionsetup{justification=raggedright,singlelinecheck=false}
	\caption{Schematic representation of the processes that contribute to the third order perturnation theory. Schetches represent the following states, beginning from the left: $\vert\phi_0\rangle,  \vert\phi_1\rangle$, and $\vert\phi_2\rangle$. 	
	}
	\label{fig7}
\end{figure}

In the weak coupling regime  where $t_\mat{el}\ll g,t_\mat{ph}\ll \omega_0\ll U$ we  use the perturbation theory where small parameters are  $g$ and $t_\mat{ph}$ and we neglect the contribution of $t_\mat{el}$. In Fig.~\ref{fig7} we show processes where the two electrons on neighbouring sites exchange a single phonon excitation. Such processes  can be described in the subspace of three functions: $\vert\phi_0\rangle = c^\dagger_{1,\uparrow} c^\dagger_{2,\downarrow}\vert\emptyset\rangle$,
$\vert\phi_1\rangle = b^\dagger_1c^\dagger_{1,\uparrow} c^\dagger_{2,\downarrow}\vert\emptyset\rangle$, and
$\vert\phi_2\rangle = b^\dagger_2c^\dagger_{1,\uparrow} c^\dagger_{2,\downarrow}\vert\emptyset\rangle$. 
We obtain the binding energy $\Delta E = E_1-E_{l\not = 1,0}$ by comparing the energy of a state when two electrons are on neighbouring sites $(l=1)$ as shown in Fig.~\ref{fig7}  with any of the states when they are further apart $(l>1)$. The difference appears in the third order of perturbation theory 
\begin{equation}
	E_1^{(3)}=\sum_{i,j=1,i \not =  j}^2{\langle\phi_0\vert H_g\vert\phi_i\rangle\langle\phi_i\vert H_{t_\mat{ph}}\vert\phi_j\rangle\langle\phi_j\vert H_g\vert\phi_0\rangle\over (\epsilon_0-\epsilon_i)(\epsilon_0-\epsilon_j)},
\end{equation} 
where $H_g$ and $H_{t_\mat{ph}}$ represent the second and the third term in Eq.~\ref{ham},  while $\epsilon_0=0$, and $\epsilon_{1}=\epsilon_{2}=\omega_0$. Note that $E_{l>1}^{(3)}=0$, which yields the binding energy
\begin{equation}
	\Delta E = 2 t_\mat{ph} {g^2\over \omega_0^2}.
	\label{deltaweak}
\end{equation}
The result is in the lowest order of perturbation theory identical to the strong coupling result in Eq.~\ref{deltast}. In both limiting cases the bipolaron is bound only when $\Delta E<0$, thus for $t_\mat{ph}<0$. 

\section{Conclusions}
Using an efficient variational method defined on the  one-dimensional chain we computed  the ground state  properties of the bipolaron  in the framework of the Holstein Hubbard model  in the presence of dispersive quantum optical phonons.  Our primary focus was on the interplay between the phonon dispersion and the Coulomb repulsion and its effects on the bipolaron effective mass, the binding energy, and the phase diagram. The main  result of our research is centerred  around the sign of the optical phonon dispersion $t_\mat{ph}$, which  takes on  a decisive role on the size of the bipolaron  and its binding energy in the presence of a strong Coulomb repulsion $U$.   In particular, when $t_\mat{ph}<0$, the signs of the electron and phonon dispersion curvatures match, and the bipolaron remains bound in the strong coupling limit even when  $U\to \infty$. At  moderate electron-coupling, a light bipolaron exists up to large values of $U$. Introducing longer range of EP coupling as well as the Coulomb interaction leads to further decrease of the effective mass, while the binding energy remains rather insensitive upon the introduced longer range interactions.

Despite a notable reduction in the effective mass within the $S1$ and $S2$ regimes, the anticipated $T_c$ does not show a detectable  increase. Our analysis suggests that  this trend results from the interplay between the effective mass and the size of the bipolaron. It appears that the decrease in the effective mass does not compensate for an increase in the size of the bipolaron. Hence, according to Eq.~\ref{Tc},  $T_c$ peaks in the $S0$ parametric range where the phonon dispersion has a negligible impact. To summarize,   increasing phonon dispersion does not lead to increased $T_c$ based on  bipolaronic condensation.

Finally, we have provided an intuitive explanation of the role of $t_\mat{ph}$ on the bipolaron binding energy. In the strong coupling regime and large $U\gg t_\mat{el}$ as well as in the weak coupling regime where $t_\mat{el}\ll g,t_\mat{ph}\ll \omega_0\ll U$,  the bipolaron binding energy takes a very simple form $\Delta E= 2t_\mat{ph}g^2/\omega_0^2$. Note also, that in this limit the singlet and the triplet bipolaron state become degenerate. Moreover, the binding emanates from the exchange of phonons between two electrons residing on adjacent sites which leads to enhanced stability of S1 and S2 bipolarons at elevated Coulomb repulsion. 



\begin{acknowledgments}
	J.B. and K.K. acknowledge the support by the program No. P1-0044 of the Slovenian Research Agency (ARIS). J.B.  acknowledge discussions with S.A. Trugman, A. Saxena and support from  the Center for Integrated Nanotechnologies, a U.S. Department of Energy, Office of Basic Energy Sciences user facility and Physics of Condensed Matter and Complex Systems Group (T-4) at Los Alamos National Laboratory. 
\end{acknowledgments}

\appendix

\section*{Appendix A: Evolution of phase diagram with electron-phonon coupling strength}
\begin{figure}[!tbh]
	\includegraphics[width=0.8\columnwidth]{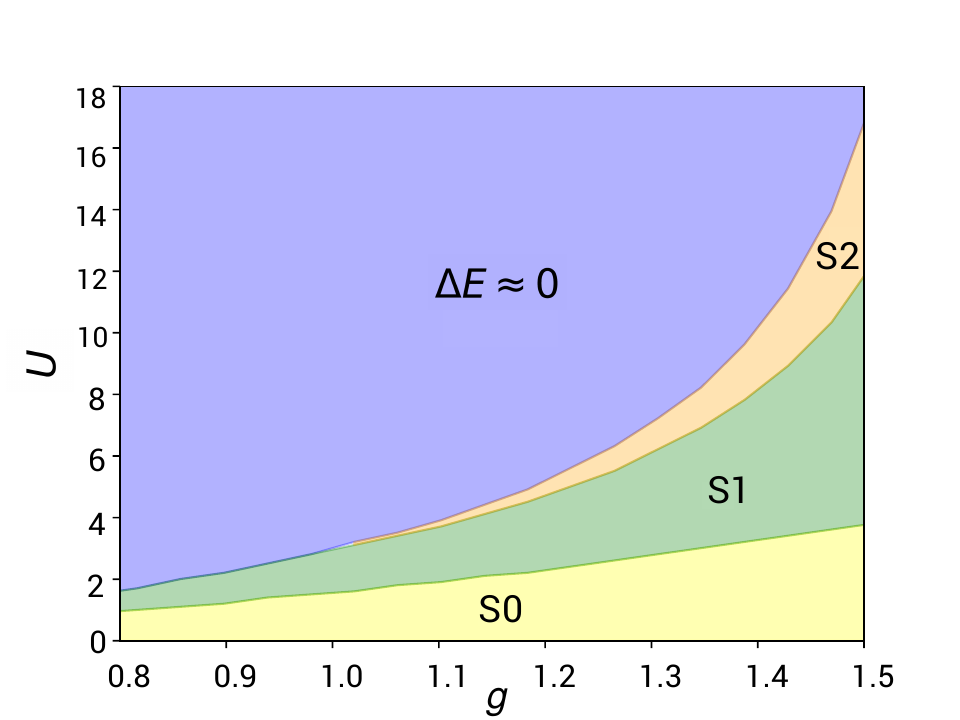}
	\captionsetup{justification=raggedright,singlelinecheck=false}
	\caption{Phase diagram ($U$, $g$) of a system with $t_\mathrm{ph} = -0.125$.}
	\label{figA1}
\end{figure}

As highlighted in the main text, the magnitude of electron-phonon coupling is a critical factor in determining whether the influence of phonon dispersion will be significant or not. This is clearly illustrated in Fig. \ref{figA1}, where a sufficiently large value of $g$ is required for the stability of the $S2$ bipolaron.

\section*{Appendix B: Finite size effects}
\begin{figure}[!tbh]
	\includegraphics[width=0.8\columnwidth]{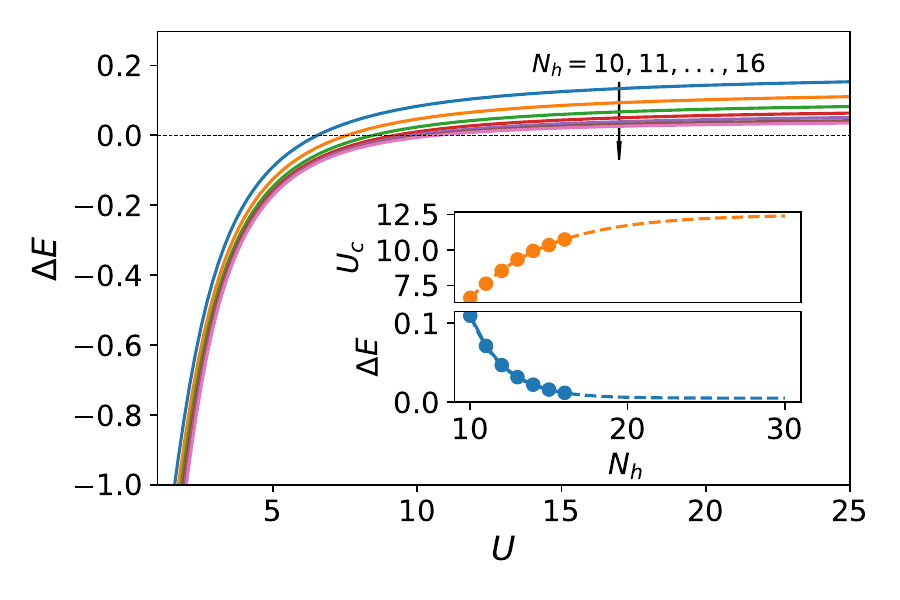}
	\captionsetup{justification=raggedright,singlelinecheck=false}
	\caption{$\Delta E$ plotted against the Hubbard parameter $U$ for  $g = \sqrt{2}, t_\mat{ph}=-0.125, \omega_0=1$ for various values of $N_h$. The direction of the arrow indicates the corresponding curve for each $N_h$. The inset illustrates the exponential approach of $U_c$ towards $U_c(N_h\to\infty)\sim 12.5 $ in the large-$N_h$ limit as well as  the decrease of $\Delta E$ towards 0 computed at  $U_c(N_h\to\infty)$ .}
	\label{figB1}
\end{figure}

In Figure \ref{figB1}, we depict $\Delta E$ as a function of the Hubbard parameter $U$ for a system with $g = \sqrt{2}$,  $\omega_0=1$ and various system sizes, characterized by $N_h$. The transition point $U_c$, marking the boundary between the bound  bipolaron and the two separate polarons, increases systematically with increasing $N_h$, ultimately converging to a constant value. Simultaneously, within the regime where $U > U_c$, $\Delta E$ decreases towards the theoretically expected value of 0.  

\vfill

\bibliography{manuaomm}
\end{document}